\documentstyle[preprint,tighten,aps]{revtex}
\begin{document}

\def\beqar{\begin{eqnarray}}
\def\eeqar{\end{eqnarray}}
\def\be{\begin{eqnarray}}
\def\ee{\end{eqnarray}}
\def\beqast{\begin{eqnarray*}}
\def\eeqast{\end{eqnarray*}}
\def\be{\begin{enumerate}}
\def\ee{\end{enumerate}}
\def\lag{\langle}
\def\rag{\rangle}
\def\fnote#1#2{\begingroup\def\thefootnote{#1}\footnote{#2}
\addtocounter{footnote}{-1}\endgroup}
\def\beq{\begin{equation}}
\def\eeq{\end{equation}}
\def\haf{\frac{1}{2}}
\def\pa{\partial}
\def\ca{{\cal A}}
\def\cb{{\cal B}}
\def\bB{\bar{B}}
\def\cH{{\cal H}_{\rm eff}}
\def\cN{{\cal N}}
\def\plb#1#2#3#4{#1, Phys. Lett. {\bf B#2}, #3 (#4)}
\def\npb#1#2#3#4{#1, Nucl. Phys. {\bf B#2}, #3 (#4)}
\def\prd#1#2#3#4{#1, Phys. Rev. {\bf D#2}, #3 (#4)}
\def\prl#1#2#3#4{#1, Phys. Rev. Lett. {\bf #2}, #3 (#4)}
\def\mpl#1#2#3#4{#1, Mod. Phys. Lett. {\bf A#2}, #3 (#4)}
\def\rep#1#2#3#4{#1, Phys. Rep. {\bf #2}, #3 (#4)}
\def\llp#1#2{\lambda_{#1}\lambda'_{#2}}
\def\lplp#1#2{\lambda'_{#1}\lambda'_{#2}}
\def\lplps#1#2{\lambda'_{#1}\lambda'^*_{#2}}
\def\lpplpp#1#2{\lambda''_{#1}\lambda''_{#2}}
\def\lpplpps#1#2{\lambda''_{#1}\lambda''^*_{#2}}
\def\slash#1{#1\!\!\!\!\!/}
\def\rpv{\slash{R_p}~}
\def\rb{B\!\!\!\!/}
\def\rl{L\!\!\!/}
\def\ckm#1#2{V_{#1} V_{#2}^*}
\def\cur#1#2{\bar{#1} #2_{-}}


\draft
\preprint{
\begin{tabular}{r}
KAIST-TH 98/12
\\
hep-ph/9808406
\end{tabular}
}
\title{
$CP$ Violating $B$ Decays with $R$-parity Violation 
}
\author{
Ji-Ho Jang 
\thanks{E-mail: jhjang@chep6.kaist.ac.kr}
and
Jae Sik Lee
\thanks{E-mail: jslee@chep6.kaist.ac.kr}
}
\address{
Department of Physics, Korea Advanced Institute of Science and
Technology \\
Taejon 305-701, Korea \\
}

\maketitle

\begin{abstract}
We study $CP$ violating $B$ decays 
in the minimal supersymmetric standard model with $R$-parity violation.
We estimate how much $R$-parity violation
modifies the SM predictions for $CP$
asymmetries in $B$ decays within the present bounds. The
effects of $R$-parity-  and lepton-number-violating couplings
on the ratio of the decay amplitude due to $R$-parity violation
to that of the SM can differ by one or two orders of magnitudes depending on
the models of the left-handed quark mixing.
It is possible to disentangle 
the $R$-parity violating effects from those of the SM and
$R$-parity-conserving supersymmetric models 
within the present bounds comparing different $CP$ violating decay amplitudes.
We also study the effects of $R$-parity- and baryon-number-violating
couplings and find that the effects could be large.
\end{abstract}


\newpage
\section{introduction}
In the upcoming experiments at $B$ factories, the large data samples will be
acquired \cite{Bfac}. One of the most important objects of these experiments
is a search for $CP$ violation in $B$ decays.
The large data on $B$ meson will enable us to probe the
physics beyond the standard model (SM) via $CP$ violating $B$ decays.
In a supersymmetric extension of the SM, there are many potential sources for
$CP$ violation in addition to the SM CKM phase. So, the SM predictions on $CP$
asymmetries in $B$ decays can be modified. 
Nondiagonality of the sfermion mass matrices in a basis where all the
couplings of neutral gauginos to fermions and sfermions are flavor diagonal
can change the SM predictions on $CP$ violation \cite{susycpv}.
The SM predictions can also be modified by the so-called $R$-parity-violating
terms.

In supersymmetric models, there are gauge invariant interactions which
violate the baryon number $B$ and the lepton number $L$ generically. To prevent
presence of these $B$ and $L$ violating interactions in supersymmetric
models, an additional global symmetry is required. This requirement leads to
the consideration of the so called $R$-parity.
The $R$-parity is given by the relation $R_p=(-1)^{(3B+L+2S)}$ where 
$S$ is  the intrinsic spin of a field.
Even though the requirement of $R_p$ conservation gives a theory
consistent with
present experimental searches, there is no good theoretical justification
for this requirement. Therefore
models with explicit $R_p$ violation ($\slash{R_p}~$) 
have been considered by many authors \cite{ago}. 

In this paper, we wish to study $CP$ violating $B$ decays 
in the minimal supersymmetric standard model (MSSM) with $\rpv$. 
We investigate how much $\rpv$ modifies the SM predictions for $CP$
asymmetries in $B$ decays within the present bounds. We emphasize that the
effects of $R_p$  and $L$ violation 
on the ratio of the decay amplitude due to $\rpv$ to that of the SM
can differ by one or two orders of magnitudes depending on
the models of the left-handed quark mixing. 
We also study the effects of $R_p$  and $B$ violation.

In the MSSM
the most general $R_p$ violating superpotential is given by
\beq
W_{R\!\!\!\!/_p}=\lambda_{ijk}L_iL_jE_k^c+\lambda_{ijk}'L_iQ_jD_k^c+
\lambda_{ijk}''U_i^cD_j^cD_k^c.
\eeq
Here $i,j,k$ are generation indices and we assume that possible bilinear terms
$\mu_i L_i H_2$ can be rotated away.
$L_i$ and $Q_i$ are the $SU(2)$-doublet lepton and the quark superfields and
$E_i^c,U_i^c,D_i^c$ are the singlet superfields respectively. 
$\lambda_{ijk}$ and
$\lambda_{ijk}''$ are antisymmetric under the interchange of the first two and
the last two generation indices respectively; $\lambda_{ijk}=-\lambda_{jik}$ and
$\lambda_{ijk}''=-\lambda_{ikj}''$. So the number of couplings is 45 (9 of the
$\lambda$ type, 27 of the $\lambda'$ type and 9 of the $\lambda''$ type).
Among these 45 couplings, 36 couplings are related with the lepton
flavor violation.

There are upper bounds on a {\it single} $L$- and $R_p$-violating couplings from
several different sources \cite{han,beta,numass,agagra,Zdecay}.
Among these, upper bounds from atomic parity violation and $eD$ asymmetry
\cite{han}, $\nu_{\mu}$ deep-inelastic scattering \cite{han},
neutrinoless double beta decay \cite{beta}, $\nu$ mass \cite{numass}, 
$K^+,t-$quark decays \cite{agagra,choroy}, and $Z$ decay width \cite{Zdecay}
 are strong.
Neutrinoless double beta decay gives
$\lambda'_{111}<3.5\times 10^{-4}$.
The bounds from $\nu$ mass are $\lambda_{133}<3\times 10^{-3}$
and $\lambda'_{133}<7\times 10^{-4}$. 

There are strong bounds on $\lambda'_{ijk}<0.012$ for $j=1$ and 2 
from $K^+$-meson decays. But, these single bounds depend on 
the models of the left-handed quark mixing.
The CKM matrix consists of the product of the mixing matrices of
the left-handed up- and down-type quarks
and we don't know the mixings of the up- and down-type quarks separately. 
Therefore, in this case,
we need some assumptions about the mixings of the left-handed
quarks to derive a single bound on $\lambda'$ coupling
from the physical process. The bounds of
$\lambda'_{i(1,2)k}<0.012$ are valid only when the mixing of the
down-type quarks dominates the CKM matrix. 
On the contrary, if the mixing of the up-type quarks dominates the CKM matrix,
the bounds on $\lambda'_{i(1,2)k}$ are totally invalid.
In general case where the CKM matrix has contributions from the up-quark
sector as well as down-quark sector, the bounds from $K^+$-meson decays 
become invalid and the typical bounds on $\lambda'_{ijk}$ with $j=2,3$ and
$\lambda'_{123,132}$ are ${\cal O}(0.1)$. We consider the general case as
well as the case in which the single bounds from $K^+$-meson decays are valid.
We find that the effects of $R_p$ violation 
can differ by one or two orders of magnitudes depending on
the models of the left-handed quark mixing.

The upper bounds on $B$- and $R_p$-violating couplings are ${\cal O}(1)$ except
$\lambda''_{112}<10^{-6}$ and $\lambda''_{113}<10^{-4}$ 
from the double nucleon decay
and $n-\bar{n}$ oscillation respectively.

In this paper we assume that
all masses of scalar partners which mediate the processes are 
100 GeV.
Extensive reviews of the limits on a single $R_p$ violating
couplings can be found in \cite{bha}\footnote{The single bounds on
$\lambda'_{132}$, $\lambda'_{232}$, and $\lambda'_{233}$ should be replaced
with 0.16 which are stronger bounds
coming from Ref.~\cite{sem}.}.

There are more stringent bounds on some products of
the $R_p$ violating couplings from 
the mixings of the neutral $K$- and $B$- mesons and
the rare leptonic decays of the $K_L$-meson, 
the muon and the tau \cite{choroy}, 
$b\bar{b}$ productions at LEP \cite{Feng},
the rare leptonic and semileptonic $B^0$ decays \cite{Lee,sem,bsll},
muon(ium) conversion, and $\tau$ and $\pi^0$ decays \cite{Ko}.


The $CP$ violating decays of $B$-meson can be induced by
the baryon number violating couplings as well as by 
the lepton number violating ones.
But, the baryon number and the lepton number 
violating couplings can not coexist in order to avoid too fast proton decays.
So we will consider the baryon number violating case
and the lepton number violating one separately.

About the baryon number violating coupling, there is a
very strong upper bound
on $\lambda''_{112}<10^{-15}$ from the proton decay 
in gauge-mediated supersymmetry breaking models
independently of the lepton number violating couplings \cite{choi}.
Recently, the study of one-loop structure of the proton decay
into very light gravitino or axino shows that all the baryon number 
violating couplings
are constrained as $\lambda''_{\rm any}<10^{-6}$ even though these bounds
depend on the precise value of the gravitino mass or 
the scale of spontaneous $U(1)_{PQ}$ breaking \cite{hwang}.

This paper is organized as follows. 
In section II, 
we introduce the general formalism for the $CP$ asymmetry in the case where the
decay amplitude contains contributions from two terms.
In section III, we consider the 
effects of $R_p$- and lepton-number-violating couplings
on the $CP$ asymmetries of neutral $B$-meson. And the
effects of $R_p$- and baryon-number-violating couplings on the CP asymmetries
are considered in section IV.
We conclude in section V.

\section{general formalism}
The time dependent $CP$ asymmetry is defined as
\beq
a_{f_{CP}}(t)\equiv\frac
{\Gamma[B^0(t)\rightarrow f_{CP}]-
\Gamma[\bar{B}^0(t)\rightarrow f_{CP}]}
{\Gamma[B^0(t)\rightarrow f_{CP}]+
\Gamma[\bar{B}^0(t)\rightarrow f_{CP}]},
\eeq
where 
$f_{CP}$ denotes the $CP$ eigenstates into which the neutral $B$ meson
decay, and
$B^0(t)$ and $\bar{B}^0(t)$ are the states that were tagged as pure
$B_d$ and $\bar{B}_d$ at the production. This $CP$ asymmetry can be rewritten by
\beq
a_{f_{CP}}(t)=a_{f_{CP}}^{\cos}\cos(\Delta M t)+
a_{f_{CP}}^{\sin}\sin(\Delta M t),
\eeq
where $\Delta M$ is the mass difference between the two physical states, and
\beq
a_{f_{CP}}^{\cos}=\frac{1-|\lambda|^2}{1+|\lambda|^2};~~~
a_{f_{CP}}^{\sin}=-\frac{2{\rm Im\lambda}}{1+|\lambda|^2}.
\eeq
Here $\lambda$ is given by 
\beqar
&&\lambda=\sqrt{\frac{<\bar{B}^0|\cH|B^0>}{<B^0|\cH|\bar{B}^0>}}
\frac{<f_{CP}|\cH|\bar{B}^0>}{<f_{CP}|\cH|B^0>}
\equiv e^{-2 i \phi_M} \frac{\bar{A}}{A}, \nonumber \\
&&<B^0|\cH|\bar{B}^0>\equiv M_{12}-\frac{i}{2} \Gamma_{12}
=\left|M_{12}-\frac{i}{2} \Gamma_{12}\right|e^{2i\phi_M},
\eeqar
using $M_{12}\gg\Gamma_{12}$.

New Physics (NP) modifies the SM predictions on both $\phi_M$ and $A$.
NP affects $B$ -- $\bar{B}$ mixing phase as follows
\beqar
\phi_M&=&\phi_M^{\rm SM}+\delta\phi_M, \nonumber \\
\delta\phi_M&=&\frac{1}{2}
\arctan\left(\frac{r_M\sin2(\phi_M^{\rm NP}-\phi_M^{\rm SM})}
                  {1+r_M\cos2(\phi_M^{\rm NP}-\phi_M^{\rm SM})}\right),
\eeqar
where 
$\phi_M^{\rm NP}$ and $\phi_M^{\rm SM}$ are defined by
\beq
<B^0|\cH^{\rm full}|\bar{B}^0>=
|M^{\rm SM}_{12}| e^{2i\phi_M^{\rm SM}}
\left(1+r_M e^{2i(\phi_M^{\rm NP}-\phi_M^{\rm SM})}\right),
\eeq
where
$r_M\equiv|M^{\rm NP}_{12}| /|M^{\rm SM}_{12}|$ and  
$M^{\rm NP}_{12}\gg\Gamma^{\rm NP}_{12}$ is assumed.
For $r_M\ll 1$, $\delta\phi_M\le r_M/2$. However for $r_M\ge 1$, 
$\delta\phi_M$ can take any value. 
In the SM, the mixing phase $\phi_M^{\rm SM}$ is $\beta$ and 0 
for $B_d$ -- $\bar{B}_d$ and $B_s$ -- $\bar{B}_s$, respectively.


If NP contributions to $A$ are dominated by one term and 
the size of the contribution is larger than that
of the sub-leading SM corrections, $A$  can be written as follows
\beq
A=A_{\rm SM} e^{i\phi_1}e^{i\delta_1}+ A_{\rm NP} e^{i\phi_2}e^{i\delta_2},~~~~
\bar{A}=A_{\rm SM} e^{-i\phi_1}e^{i\delta_1}+ 
A_{\rm NP} e^{-i\phi_2}e^{i\delta_2},
\eeq
where $A_{\rm SM,NP}$ are real magnitudes, 
$\phi_{1,2}$ are $CP$ violating phases
and $\delta_{1,2}$ are $CP$ conserving phases. 
For the sizes of the sub-leading SM corrections and the contributions of the
$R_p$-conserving supersymmetric model, see Ref. \cite{worah}.

With $\phi_{12}=\phi_1-\phi_2$,  
$\delta_{12}=\delta_1-\delta_2$, and $r_D \equiv A_{\rm NP}/A_{\rm SM}$,
\beqar
a_{f_{CP}}^{\cos}&=&
-\frac{2r_D\sin\phi_{12}\sin\delta_{12}}{1+2r_D\sin\phi_{12}\sin\delta_{12}} 
\approx -2r_D\sin\phi_{12}\sin\delta_{12},
\nonumber \\
a_{f_{CP}}^{\sin}&=&
\frac{\sin(2\phi_M+\phi_1)-2r_D\sin\phi_{12}\cos(2\phi_M+2\phi_1+\delta_{12})}
{1+2r_D\sin\phi_{12}\sin\delta_{12}}\nonumber \\ 
&\approx &
\sin2(\phi_M+\phi_1)-2r_D\sin\phi_{12}\cos2(\phi_M+\phi_1)\cos\delta_{12},
\eeqar
to the first order in $r_D$.

For the rest of this paper, we concentrate on $a_{f_{CP}}^{\sin}$.
To this end we write
\beqar
a_{f_{CP}}^{\sin}\equiv
\sin2(\phi_M+\phi_1+\delta\phi_D)\equiv
\sin2\phi,
\eeqar
For $r_D\ll 1$, $\delta\phi_D\le r_D$. However for $r_D\ge 1$, 
$\delta\phi_D$ can take any value. 
In the following two sections, we will calculate $r_D$ 
for several $CP$ violating decay modes. 

Note that NP contribution to the mixing phase $\phi_M$ is
universal for all kinds of decay modes. 
So, one can identify NP contributions to
$CP$ violating $B$ decays independently of the NP contribution to the
mixing by
considering two different decay modes simultaneously.

\section{$R_p$ and $L$ violation} 
In this section, we consider the effects of $R_p$ and the lepton number
violating couplings ($\lambda'$)
assuming the baryon number violating couplings $\lambda''$'s vanish.

Firstly we assume
$V_{\rm CKM}$ is given by only down-type quark sector mixing.
In this case, $r_M$ and $r_D(B_d\rightarrow \psi K_S,\phi K_S)$ are estimated
in Ref. \cite{guetta} as follows,
\beqar
&&r_M(B_d) \simeq 10^8 ~|\lplp{n13}{n31}|\left(\frac{100~{\rm GeV}}{
M_{\tilde\nu}}\right)^2, \nonumber \\
&&r_D(B_d\rightarrow \psi K_S)< 0.02,\nonumber \\
&&r_D(B_d\rightarrow \phi K_S)< 0.8, 
\eeqar
and $|\phi(B_d\rightarrow \psi K_S) -\phi(B_d\rightarrow \phi K_S)| 
< {\cal O}(1)$.
$r_M(B_s)$ is given by replacing $|\lplp{n13}{n31}|$ with
$|\lplp{n23}{n32}|$ in $r_M(B_d)$.
In this section, we wish to investigate other decay modes and discuss
how much the effects of $\rpv$ differ depending on the models of the
left-handed quark mixings.

From Eq.~(1), we obtain the following four-fermion
effective Lagrangian due to the exchange of the sleptons 
\beqar
{\cal L}^{\rm eff,2u-2d}_{\rpv}&=&
\frac{4G_F}{\sqrt{2}}
{\cal C}_{ijkl}^{\rl}
(\bar{d}_i P_L u_j)(\bar{u}_k  P_R d_l),
\nonumber \\
{\cal L}^{\rm eff,4d}_{\rpv}&=&
\frac{4G_F}{\sqrt{2}}
{\cal N}_{ijkl}^{\rl}
(\bar{d}_i P_L d_j)(\bar{d}_k  P_R d_l),
\eeqar
where $P_{L,R}=(1\mp\gamma_5)/2$ and the dimensionless couplings
${\cal C}_{ijkl}^{\rl}$ and ${\cal N}_{ijkl}^{\rl}$ are given by
\beqar
{\cal C}_{ijkl}^{\rl}&=&
\frac{\sqrt{2}}{4G_F}
\sum_{n,p,q=1}^3 \frac{1}{M^2_{\tilde{l}_n}}
V_{kq}V^*_{jp}\lambda'_{npi} \lambda'^*_{nql},
\nonumber \\
{\cal N}_{ijkl}^{\rl}&=&
\frac{\sqrt{2}}{4G_F}
\sum_{n=1}^3 \frac{1}{M^2_{\tilde{l}_n}}
\lambda'_{nji} \lambda'^*_{nkl}.
\eeqar
From the above effective Lagrangian, we calculate the amplitudes $A$ for the
several decay modes under the factorization assumption and the results are
shown in the Appendix.

In Table \ref{haha1}, we show the R-parity- and lepton-number-violating
product combinations which significantly contribute to each process
assuming $V_{\rm CKM}$ is given by only down-type quark sector mixing.
For the decay mode $B_d\rightarrow \psi K_S$, there are four kinds of
competitive contributions and the most significant one comes from 
$\lplp{332}{333}$ within present bounds
\footnote{ In Ref. \cite{guetta}, only the contributions
from $\lplp{n22}{n23}$ are considered.}.
Typically, the constraints are order of $10^{-4}$ or $10^{-3}$.
The decay modes with $10^{-3}$ constraint are
$B_d\rightarrow \phi K_S$, $B_d\rightarrow \pi^0 K_S$, 
$B_s\rightarrow \phi K_S$, $B_d\rightarrow \phi \pi^0$, 
and $B_d\rightarrow \pi^0\pi^0$. 
So these five decay modes are important
ones in the presence of $R_P$ violation. See Table \ref{haha2} for the
estimated values of $r_D$. 

The supersymmetric contributions
to the decay modes $B_d\rightarrow \phi K_S$ and
$B_d\rightarrow \pi^0 K_S$ are not dominated by only $\rpv$ since
there are comparable contributions from nondiagonal sfermion mass matrices
to these decay modes, see the second paper of Ref. \cite{worah}.
And the upcoming $B$ experiments will initially take data 
at $\Upsilon(4s)$ where only the $B_d$ can be studied and 
the mode $B_d\rightarrow \pi^0\pi^0$ suffers from the large SM uncertainties. 
For the decay mode $B_d\rightarrow \phi \pi^0$,
the SM prediction for the branching ratio of this decay mode is quite small :
${\cal B}_{\rm SM}(B_d\rightarrow \phi \pi^0)=1.9\times 10^{-8}$ \cite{ali}.
Consequently, it would be hard to measure $CP$ violation 
considering only one decay mode unless $\rpv$ enhance the
branching ratio of this mode significantly. But, 
the $R_p$- and $L$-violating effects can be disentangled
from those of the SM or $R_p$-conserving supersymmetric models if we
compare two or more decay modes. For example, let's think
about the decay modes of $B_d\rightarrow \psi K_S$ and
$B_d\rightarrow \phi K_S$. 
The difference between $CP$ violating phases of
these two decay modes vanishes 
in the SM or $R_p$-conserving supersymmetric models.
But, it does not vanish in the $R_p$-violating model.

Now, let's think the general case in which
the down-type quark mixing does not dominates $V_{\rm CKM}$.
In this case, the strong bounds $|\lambda'_{ijk}|<0.012$ with $j=1,2$
from $K^+$-meson decays becomes invalid. In this case,
the typical bounds on $\lambda'_{ijk}$ with $i=2,3$ are ${\cal O}(0.1)$.
This means that the constraints given in Table \ref{haha1} can become weaker
by one or two orders of magnitudes.
For example, let us consider the
contribution of $\lplp{222}{223}$ to the $CP$ asymmetry in the mode
$B_d\rightarrow \psi K_S$. Neglecting the constraint from $K^+$ decays, the
constraint on this combination is $3.2\times 10^{-2}$ from $D$-decay
\cite{bha}. 
Using this constraint, one can
obtain $r_D(B_d\rightarrow \psi K_S)=7.5$. 
Similarly, we find that the typical size of
$r_D$ of {\it all} decay modes is ${\cal O}(1)$
if we neglect the constraint from $K^+$ decays. 
It means that it is possible to disentangle 
the $R$-parity violating effects from those of the SM and
$R$-parity-conserving supersymmetric models.
In this case, one can also identify the NP effects independently of the NP
contributions to the mixing by
taking account of the differences between the angles $\phi$'s
of the first five modes in Table \ref{haha2}.

\section{$R_p$ and $B$ violation} 
In this section, we consider the effects of $R_p$ and the baryon number
violating couplings ($\lambda''$)
assuming the lepton number violating couplings $\lambda'$'s vanish.

From Eq. (1), we obtain the following four-fermion
effective Lagrangian due to the exchange of the squarks
\beqar
{\cal L}^{\rm eff,2u-2d}_{\rpv}&=&
\frac{4G_F}{\sqrt{2}}
{\cal C}_{ijkl}^{\rb}\left[
(\bar{u}_i \gamma^{\mu}P_R u_j)(\bar{d}_k \gamma_{\mu} P_R d_l)-
(\bar{d}_k \gamma^{\mu}P_R u_j)(\bar{u}_i \gamma_{\mu} P_R d_l)\right],
\nonumber \\
{\cal L}^{\rm eff,4d}_{\rpv}&=&
\frac{4G_F}{\sqrt{2}}
{\cal N}_{ijkl}^{\rb}
(\bar{d}_i \gamma^{\mu}P_R d_j)(\bar{d}_k \gamma_{\mu} P_R d_l),
\eeqar
where $P_{L,R}=(1\mp\gamma_5)/2$ and the dimensionless couplings
${\cal C}_{ijkl}^{\rb}$ and ${\cal N}_{ijkl}^{\rb}$ are given by
\beqar
{\cal C}_{ijkl}^{\rb}&=&
\frac{\sqrt{2}}{4G_F}
\sum_{n=1}^3 \frac{2}{M^2_{\tilde{d}_n}}
\lambda''_{ikn} \lambda''^*_{jln},
\nonumber \\
{\cal N}_{ijkl}^{\rb}&=&
\frac{\sqrt{2}}{4G_F}
\sum_{n=1}^3 \frac{1}{M^2_{\tilde{u}_n}}
\lambda''_{nik} \lambda''^*_{njl}.
\eeqar
From the above effective Lagrangian, we calculate the amplitudes for 
several decay modes using the factorization assumption and the results are
shown in the Appendix.

By the inspection of ${\cal N}_{ijkl}^{\rb}$, one can easily see that
 $R_p$- and $B$-violating
couplings does not contribute $B$ -- $\bar{B}$ mixing 
and $B_d\rightarrow\phi K_S$
since $\lambda''_{ijk}$
is antisymmetric under the exchange of the last two indices.

The present bounds on $\lambda''$ are so poor that $r_D$'s are
generally quite large except $B_d\rightarrow\pi\pi$ mode
: see Table \ref{haha3}. 
Large $r_D$ means two things.
One thing is that it is
possible to have large $CP$ violation completely different
from the SM predictions. 
The other thing is that one can obtain more stringent bounds on
the product combinations 
if the measured branching ratios of the decay modes are 
consistent with the SM predictions \cite{jl}.

Note that one product combination contributes to two and more decay modes, see
Table \ref{haha3}. 
In this case, the differences of $CP$ phases $\phi$'s
of the decay modes are exactly the same as that of the SM. 

In gauge-mediated supersymmetry breaking models, $\lambda''$ are severely
constrained from the proton decay \cite{choi,hwang}. So, the contributions of
$R_p$- and $B$-violating couplings to $CP$ violating $B$ decays can be
safely ignored.

\section{Conclusion} 
To conclude, we study $CP$ violating $B$ decays 
in the minimal supersymmetric standard model with $\rpv$. 
We estimate how much $\rpv$ modifies the SM predictions for $CP$
asymmetries in $B$ decays within the present bounds. The
effects of $R_p$  and $L$ violation
on the ratio of the decay amplitude due to $\rpv$
to that of the SM
can differ by one or two orders of magnitudes depending on
the models of the left-handed quark mixing.
It is possible to disentangle 
the $R$-parity violating effects from those of the SM and
$R$-parity-conserving supersymmetric models 
within the present bounds.
We also study the effects of $R_p$ and $B$ violation 
and find that the effects could be large or the contributing product
combinations can be strongly constrained by the near future experiments on $B$
mesons.  The effects of $R_p$ and $B$ violation 
can be ignored in gauge-mediated supersymmetric models.

\section*{acknowledgements}
We thank Y. G. Kim and P. Ko for their helpful remarks.
This work was supported in part by KAIST Basic Science Research Program
(J.S.L.).

\section*{Appendix}
In this appendix, we present all decay amplitudes relevant to our analysis. 
We don't need to know the exact values of the form factor since they are
irrelevant in the calculation of $r_D$ for the most of cases.  
For the numerical calculation, we use the following values for the quark
masses :
$m_u =$ 4.2 MeV, $m_d =$ 7.6 MeV, $m_s =$ 122 MeV, $m_c =$ 1.3 GeV,
$m_b =$ 4.88 GeV and we take $N =$3.
\subsection{SM}
The amplitudes in SM are calculated using the effective Hamiltonian formalism. 
The short and long distance QCD effects in the nonleptonic decays are 
separated by means of the operator product expansion. 
For the numerical values of the Wilson coefficients ( short distance effects ),
we use the the values in Ref. ~\cite{ali}. The long distance contributions
of the hadronic matrix elements are calculated under the factorization 
approximation.
\beqar
&&A(\bar{B}^0 \rightarrow \psi K_S) = 
\frac{G_F}{\sqrt{2}} \left[ \ckm{cb}{cs} a_2 -
\ckm{tb}{ts} ( a_3 + a_5 + a_7 + a_9 ) \right] 
\nonumber \\
&& ~~~~~~~~~~~~~~~~~~~~~~~~
\times < K_S | \cur{s}{b} | \bar{B}^0 > < \psi | \cur{c}{c} | 0 > 
\\
&&A(\bar{B}^0 \rightarrow \phi K_S) =
- \frac{G_F}{\sqrt{2}} \ckm{tb}{ts} \left[
a_3 + a_4 + a_5 - \frac{1}{2} ( a_7 + a_9 + a_{10} ) \right] 
\nonumber \\
&& ~~~~~~~~~~~~~~~~~~~~~~~~
\times < K_S | \cur{s}{b} | \bar{B}^0 >< \phi | \cur{s}{s} | 0 >
\\
&&A(\bar{B}^0 \rightarrow \pi^0 K_S) =
\frac{G_F}{\sqrt{2}} \left[ \left\{ \ckm{ub}{us} a_2 +
\frac{3}{2} \ckm{tb}{ts} (a_7 - a_9) \right\}
< \pi^0 | \bar{u}u_- | 0 >< K_S | \bar{s}b_- | \bar{B}^0 > \right.
\nonumber \\
&& ~~~~~~~~~~
\left. - \ckm{tb}{ts} \left\{ a_4 - \frac{1}{2} a_{10} +
\frac{m_K^2( 2 a_6 - a_8)}{(m_d+m_s)(m_b-m_d)} \right\}
< K_S | \bar{s}d_- | 0 >< \pi^0 | \bar{d}b_- | \bar{B}^0 > \right]
\\
&&A(\bar{B}^0 \rightarrow D^+ D^-) =
\frac{G_F}{\sqrt{2}} \left[ \ckm{cb}{cd} a_1 - \ckm{tb}{td}
\left\{ a_4 + a_{10} +
\frac{ 2 m_D^2 ( a_6 + a_8 )}{( m_c + m_d )( m_b - m_c )} \right\} \right]
\nonumber \\
&& ~~~~~~~~~~~~~~~~~~~~~~~~
\times < D^+ | \cur{c}{b} | \bar{B}^0 >< D^- | \cur{d}{c} | 0 >
\\
&&A(\bar{B}^0 \rightarrow D_{CP} \pi^0) =
\frac{G_F}{\sqrt{2}} ( \ckm{cb}{ud} \pm \ckm{ub}{cd} ) a_2 
< \pi^0 | \cur{d}{b} | \bar{B}^0 >< D_{CP} | \cur{c}{u} | 0 >
\\
&&A(\bar{B}^0 \rightarrow D_{CP} \rho^0) =
\frac{G_F}{\sqrt{2}} ( \ckm{cb}{ud} \pm \ckm{ub}{cd} ) a_2 
< \rho^0 | \cur{d}{b} | \bar{B}^0 >< D_{CP} | \cur{c}{u} | 0 >
\\
&&A(\bar{B}_s \rightarrow \phi K_S) =
- \frac{G_F}{\sqrt{2}} \ckm{tb}{td} \left[ a_3 + a_4 + a_5
- \frac{1}{2} \left\{ a_7 + a_9 + a_{10} \right\} +
\frac{m_{\phi}^2 ( 2 a_6 - a_8)}{ 2 m_s ( m_b - m_s)} \right]
\nonumber \\
&& ~~~~~~~~~~~~~~~~~~~~~~~~
\times < K_S | \cur{d}{b} | \bar{B}_s >< \phi| \cur{s}{s} | 0 >
\\
&&A(\bar{B}^0 \rightarrow \phi \pi^0) =
- \frac{G_F}{\sqrt{2}} \ckm{tb}{td} \left\{ a_3 + a_5 -
\frac{1}{2} ( a_7 + a_9 ) \right\}
< \pi^0 | \bar{d}b_- | \bar{B}^0 >< \phi | \bar{s}s_- | 0 >
\\
&&A(\bar{B}^0 \rightarrow \pi^+ \pi^-) =
\frac{G_F}{\sqrt{2}} \left[ \ckm{ub}{ud} a_1 -  \ckm{tb}{td}    
\left\{ a_4 + a_{10} +
\frac{ 2 m_{\pi}^2 ( a_6 + a_8 )}{( m_u + m_d )( m_b - m_u )} \right\} \right]
\nonumber \\
&& ~~~~~~~~~~~~~~~~~~~~~~~~
\times < \pi^+ | \cur{u}{b} | \bar{B}^0 >< \pi^- | \cur{d}{u} | 0 >
\\
&&A(\bar{B}^0 \rightarrow \pi^0 \pi^0) =
- \frac{2 G_F}{\sqrt{2}} \left[ \ckm{ub}{ud} a_2 + \ckm{tb}{td}
\left\{ a_4 + \frac{3}{2} ( a_7 - a_9 ) - \frac{1}{2} a_{10} 
\right.\right.
\nonumber \\
&& ~~~~~~~~~~~~~~~~~~~~~~~~
\left.\left.
+\frac{ m_{\pi}^2 ( 2 a_6 - a_8 )}{ 2 m_d ( m_b - m_u )} \right\} \right]
\times < \pi^0 | \cur{d}{b} | \bar{B}^0 >< \pi^0 | \cur{u}{u} | 0 >
\eeqar
The $\pm$ sign in $\bar{B}^0 \rightarrow D_{CP} \pi^0(\rho^0)$ decay modes
corresponds to the $CP$-even and $CP$-odd eigenstates of $D_{CP}$
and the same convention is applied to the $R_p$ violation case.
In the numerical estimation of $\bar{B}^0 \rightarrow \pi^0 K_S$ decay modes,
we assume that 
$|< \pi^0 | \bar{u}u_- | 0 >< K_S | \bar{s}b_- | \bar{B}^0 >|\approx
|< K_S | \bar{s}d_- | 0 >< \pi^0 | \bar{d}b_- | \bar{B}^0 >|$.

\subsection{$R_p$ and $L$ violation}
In this case, the running effects of the $R_p$ violating couplings are
neglected. The hadronic matrix elements are also calculated under the
factorization assumption. 

\beqar
&&A(\bar{B}^0 \rightarrow \psi K_S)  =  
\sum_{n,i,j} \frac{1}{M_{\tilde{l}_n}^2} \frac{1}{8 N} \lplps{ni2}{nj3}
\ckm{2j}{2i}
< K_S | \cur{s}{b} | \bar{B}^0 > < \psi | \cur{c}{c} | 0 > 
\\
&&A(\bar{B}^0 \rightarrow \phi K_S)  = 
\sum_{n} \frac{1}{M_{\tilde{l}_n}^2} \frac{1}{8 N} \left[
\lplps{n22}{n23} + \lplps{n32}{n22} \right]
< K_S | \cur{s}{b} | \bar{B}^0 >< \phi | \cur{s}{s} | 0 >
\\
&&A(\bar{B}^0 \rightarrow \pi^0 K_S)  = 
\sum_{n} \frac{1}{M_{\tilde{l}_n}^2} \left[
\left\{ \frac{1}{8 N} ( \sum_{i,j}\lplps{ni2}{nj3} \ckm{1j}{1i} -
\lplps{n12}{n13} + \lplps{n31}{n21} ) \right. \right.
\nonumber \\
&& ~~~~~~~~~~~~~~
+ \left. \frac{m_{\pi}^2}{8 m_d(m_b -m_s)}
(\lplps{n11}{n23} - \lplps{n32}{n11} ) \right\} 
< \pi^0 | \bar{u}u_- | 0 >< K_S | \bar{s}b_- | \bar{B}^0 > 
\nonumber \\
&& ~~~~~~~~~~~~~~
+ \left\{ \frac{1}{8 N} (\lplps{n11}{n23} - \lplps{n32}{n11} )
-\frac{m_{K_S}^2}{4 (m_b-m_d)(m_d +m_s)}
(\lplps{n12}{n13} - \lplps{n31}{n21} ) \right\}
\nonumber \\
&& ~~~~~~~~~~~~~~
\left.\times < K_S | \bar{s}d_- | 0 >< \pi^0 | \bar{d}b_- | \bar{B}^0 >
\right]
\\
&&A(\bar{B}^0 \rightarrow D^+ D^-)  = 
\sum_{n,i,j} \frac{1}{M_{\tilde{l}_n}^2} 
\frac{m_{D^-}^2}{4(m_d + m_c)(m_b - m_c)} \lplps{ni1}{nj3} \ckm{2j}{2i}
\nonumber \\
&& ~~~~~~~~~~~~~~~~~~~~~~~~
 \times < D^+ | \cur{c}{b} | \bar{B}^0 >< D^- | \cur{d}{c} | 0 >
\\
&&A(\bar{B}^0 \rightarrow D_{CP} \pi^0)  = 
\sum_{n,i,j} \frac{1}{M_{\tilde{l}_n}^2} \lplps{ni1}{nj3}
\frac{1}{8 N}
\left[ \ckm{2j}{1i} \pm \ckm{1j}{2i} \right]
< \pi^0 | \cur{d}{b} | \bar{B}^0 >< D_{CP} | \cur{c}{u} | 0 >
\\
&&A(\bar{B}^0 \rightarrow D_{CP} \rho^0)  = 
- \sum_{n,i,j} \frac{1}{M_{\tilde{l}_n}^2} \lplps{ni1}{nj3}
\frac{1}{8 N}
\left[ \ckm{2j}{1i} \pm \ckm{1j}{2i} \right]
\nonumber \\
&& ~~~~~~~~~~~~~~~~~~~~~~~~
\times < \rho^0 | \cur{d}{b} | \bar{B}^0 >< D_{CP} | \cur{c}{u} | 0 >
\\
&&A(\bar{B}_s \rightarrow \phi K_S)  = 
- \sum_{n} \frac{1}{M_{\tilde{l}_n}^2} \left[
\left\{ \frac{1}{8 N} ( \lplps{n12}{n23} + \lplps{n32}{n21} +
\lplps{n22}{n13} + \lplps{n31}{n22} ) \right. \right.
\nonumber \\
&& ~~~~~~~~~~~~~~~~~~~~~~~~
 \left. + \frac{m_{K_S}^2}{4 ( m_s + m_d )( m_s + m_b )}
( \lplps{n12}{n23} + \lplps{n32}{n21} +
\lplps{n21}{n23} + \lplps{n32}{n12} ) \right\}
\nonumber \\
&& ~~~~~~~~~~~~~~~~~~~~~~~~
\times < \phi | \cur{s}{b} | \bar{B}_s >< K_S | \cur{d}{s} | 0 >
\nonumber \\
&& ~~~~~~~~~~~~~~~~~~~~~~~~
- \left. \frac{1}{8 N} ( \lplps{n21}{n23} + \lplps{n32}{n12} )
< K_S | \cur{d}{b} | \bar{B}_s >< \phi| \cur{s}{s} | 0 > \right]
\\
&&A(\bar{B}^0 \rightarrow \phi \pi^0)  = 
\sum_{n} \frac{1}{M_{\tilde{l}_n}^2} \frac{1}{8 N} 
(\lplps{n21}{n23}+\lplps{n32}{n12})
< \pi^0 | \bar{d}b_- | \bar{B}^0 >< \phi | \bar{s}s_- | 0 >
\\
&&A(\bar{B}^0 \rightarrow \pi^+ \pi^-)  = 
- \sum_{n,i,j} \frac{1}{M_{\tilde{l}_n}^2} 
\frac{m_{\pi^-}^2}{4(m_d + m_u)(m_b - m_u)} \lplps{ni1}{nj3} \ckm{1j}{1i}
\nonumber \\
&& ~~~~~~~~~~~~~~~~~~~~~~~~
\times < \pi^+ | \cur{u}{b} | \bar{B}^0 >< \pi^- | \cur{d}{u} | 0 >
\\
&&A(\bar{B}^0 \rightarrow \pi^0 \pi^0)  = 
\sum_{n} \frac{1}{M_{\tilde{l}_n}^2}
\left[ \sum_{i,j} \frac{1}{4 N} \lplps{ni1}{nj3} \ckm{1j}{1i}
\right. \nonumber \\
&& ~~~~~~~~~~~~~~~~~~~~~~~~
- \left. \left\{ \frac{1}{4 N} - 
            \frac{m_{\pi^0}^2}{4 m_d ( m_b - m_d)} \right\}
( \lplps{n11}{n13} - \lplps{n31}{n11} ) \right]
\nonumber \\
&& ~~~~~~~~~~~~~~~~~~~~~~~~
 \times < \pi^0 | \cur{d}{b} | \bar{B}^0 >< \pi^0 | \cur{u}{u} | 0 >
\eeqar
In $\bar{B}^0 \rightarrow \pi^0 K_S$ and $\bar{B}^0 \rightarrow \phi K_S$
modes, we assume that the magnitudes of two form factors 
are approximately same.

\subsection{$R_p$ and $B$ violation}
The decay amplitudes for $R_p$ and $B$ violation are calculated in the
similar way as the case of $R_p$ and $L$ violation.
\beqar
&&A(\bar{B}^0 \rightarrow \psi K_S)  =  
- \sum_{n} \frac{1}{2 M_{\tilde{d}_n}^2} ( 1 - \frac{1}{N} )
\lpplpps{22n}{23n}
< K_S | \cur{s}{b} | \bar{B}^0 > < \psi | \cur{c}{c} | 0 > 
\\
&&A(\bar{B}^0 \rightarrow \phi K_S)  =  0
\\
&&A(\bar{B}^0 \rightarrow \pi^0 K_S)  = 
\sum_{n} \left[\left\{ \frac{1}{2 M_{\tilde{d}_n}^2} \lpplpps{12n}{13n} -
\frac{1}{2 M_{\tilde{u}_n}^2} \lpplpps{n12}{n13} \right\} ( 1 - \frac{1}{N} ) 
\right.
\nonumber \\
&& ~~~~~~~~~~~~~~~~~~~~~~~~
\times < \pi^0 | \bar{u}u_- | 0 >< K_S | \bar{s}b_- | \bar{B}^0 >
\nonumber \\
&& ~~~~~~~~~~~~~~~~~~~~~~~~
-\left.\frac{1}{2 M_{\tilde{u}_n}^2} ( 1 - \frac{1}{N} )\lpplpps{n12}{n13}
< K_S | \bar{s}d_- | 0 >< \pi^0 | \bar{d}b_- | \bar{B}^0 >
\right]
\\
&&A(\bar{B}^0 \rightarrow D^+ D^-)  = 
- \sum_{n} \frac{1}{2 M_{\tilde{d}_n}^2} ( 1 - \frac{1}{N} )
\lpplpps{21n}{23n}
< D^+ | \cur{c}{b} | \bar{B}^0 >< D^- | \cur{d}{c} | 0 >
\\
&&A(\bar{B}^0 \rightarrow D_{CP} \pi^0)  = 
\sum_{n} \frac{1}{2 M_{\tilde{d}_n}^2} ( 1 - \frac{1}{N}) \left[
\lpplpps{21n}{13n} \pm \lpplpps{11n}{23n} \right]
\nonumber \\
&& ~~~~~~~~~~~~~~~~~~~~~~~~
\times < \pi^0 | \cur{d}{b} | \bar{B}^0 >< D_{CP} | \cur{c}{u} | 0 >
\\
&&A(\bar{B}^0 \rightarrow D_{CP} \rho^0)  = 
- \sum_{n} \frac{1}{2 M_{\tilde{d}_n}^2} ( 1 - \frac{1}{N}) \left[
\lpplpps{21n}{13n} \pm \lpplpps{11n}{23n} \right]
\nonumber \\
&& ~~~~~~~~~~~~~~~~~~~~~~~~~~~
\times < \rho^0 | \cur{d}{b} | \bar{B}^0 >< D_{CP} | \cur{c}{u} | 0 >
\\
&&A(\bar{B}_s \rightarrow \phi K_S)  = 
- \sum_{n} \frac{1}{2 M_{\tilde{u}_n}^2} ( 1 - \frac{1}{N} )
\lpplpps{n12}{n23}
\nonumber \\
&& ~~~~~~~~~~~~~~~~~~~~~~~~
 \times \left[
< \phi | \cur{s}{b} | \bar{B}_s >< K_S | \cur{d}{s} | 0 > -
< K_S | \cur{d}{b} | \bar{B}_s >< \phi| \cur{s}{s} | 0 > \right]
\\
&&A(\bar{B}^0 \rightarrow \phi \pi^0)  = 
\sum_{n}\frac{1}{2 M_{\tilde{u}_n}^2}( 1 - \frac{1}{N} )\lpplpps{n12}{n23}
< \pi^0 | \bar{d}b_- | \bar{B}^0 >< \phi | \bar{s}s_- | 0 >
\\
&&A(\bar{B}^0 \rightarrow \pi^+ \pi^-)  = 
- \sum_{n} \frac{1}{2 M_{\tilde{d}_n}^2} ( 1 - \frac{1}{N} )
\lpplpps{11n}{13n} 
< \pi^+ | \cur{u}{b} | \bar{B}^0 >< \pi^- | \cur{d}{u} | 0 >
\\
&&A(\bar{B}^0 \rightarrow \pi^0 \pi^0)  = 
 \sum_{n} \frac{1}{M_{\tilde{d}_n}^2} ( 1 - \frac{1}{N} )
\lpplpps{11n}{13n}
< \pi^0 | \cur{d}{b} | \bar{B}^0 >< \pi^0 | \cur{u}{u} | 0 >
\eeqar
In $\bar{B}_s \rightarrow \phi K_S$ decay mode, we assume \cite{gatto}
\beqar
\frac{
< K_S | \cur{d}{b} | \bar{B}_s >< \phi| \cur{s}{s} | 0 > -
< \phi | \cur{s}{b} | \bar{B}_s >< K_S | \cur{d}{s} | 0 > 
}
{< K_S | \cur{d}{b} | \bar{B}_s >< \phi| \cur{s}{s} | 0 >} \approx
{\cal O}(1). \nonumber
\eeqar

\begin{table}
\caption{\label{haha1}
R-parity- and lepton-number-violating
product combinations which significantly contribute 
within present bounds \protect\cite{bha,sem,bsll} 
assuming $V_{\rm CKM}$ is given by only down-type quark sector mixing.
Constraints on the magnitudes of the product combinations are also shown.
}
\vspace{0.1 cm}
\begin{tabular}{lcc}
Decay Modes & Dominating & Constraint\\
& Combination &    \\
\hline
$\bar{B}_d \rightarrow \psi K_S$ 
& $\lplp{n22}{n23}V_{22}V_{22}$ & $1.4 \times 10^{-4}$ \\
& $\lplp{322}{333}V_{23}V_{22}$ & $2.3 \times 10^{-4}$ \\
& $\lplp{332}{323}V_{22}V_{23}$ & $2.3 \times 10^{-4}$ \\
& $\lplp{332}{333}V_{23}V_{23}$ & $3.9 \times 10^{-4}$ \\
\hline
$\bar{B}_d \rightarrow \phi K_S$ 
& $\lplp{132}{122}$ & $1.1 \times 10^{-3}$ \\
& $\lplp{232}{222}$ & $1.1 \times 10^{-3}$ \\
& $\lplp{332}{322}$ & $5.8 \times 10^{-3}$ \\
\hline
$\bar{B}_d \rightarrow \pi^0 K_S$  
& $\lplp{231}{221},~\lplp{232}{211}$ & $1.9 \times 10^{-3}$ \\
& $\lplp{331}{321},~\lplp{332}{311}$ & $5.8 \times 10^{-3}$ \\
\hline
$\bar{B}_d \rightarrow D^+ D^-$ 
& $\lplp{n21}{n23}V_{22}V_{22}$ & $1.4 \times 10^{-4}$ \\
& $\lplp{321}{333}V_{23}V_{22}$ & $2.3 \times 10^{-4}$ \\
& $\lplp{331}{323}V_{22}V_{23}$ & $2.3 \times 10^{-4}$ \\
& $\lplp{331}{333}V_{23}V_{23}$ & $3.9 \times 10^{-4}$ \\
\hline
$\bar{B}_d \rightarrow D_{CP} ~\pi^0(\rho^0)$
& $\lplp{311}{333}V_{23}V_{11}$ & $2.3 \times 10^{-4}$ \\
& $\lplp{311}{323}V_{22}V_{11}$ & $1.4 \times 10^{-4}$ \\
& $\lplp{211}{223}V_{22}V_{11}$ & $1.4 \times 10^{-4}$ \\
& $\lplp{n21}{n13}V_{11}V_{22}$ & $1.4 \times 10^{-4}$ \\
& $\lplp{331}{313}V_{11}V_{23}$ & $2.3 \times 10^{-4}$ \\
\hline
$\bar{B}_s \rightarrow \phi K_S$ 
& $\lplp{132}{121},~\lplp{132}{112}$,~$\lplp{232}{221},~\lplp{232}{212}$
,~$\lplp{231}{222}$ & $1.9 \times 10^{-3}$ \\
& $\lplp{332}{321},~\lplp{331}{322},~\lplp{332}{312}$ & $5.8 \times 10^{-3}$ \\
\hline
$\bar{B}_d \rightarrow \phi \pi^0$ 
& $\lplp{132}{112},~\lplp{232}{212}$ & $1.9 \times 10^{-3}$ \\
& $\lplp{332}{312}$ & $5.8 \times 10^{-3}$ \\
\hline
$\bar{B}_d \rightarrow \pi^+ \pi^- $ 
& $\lplp{211}{213}V_{11}V_{11}$ & $1.4 \times 10^{-4}$ \\ 
& $\lplp{311}{313}V_{11}V_{11}$ & $1.4 \times 10^{-4}$ \\
$\bar{B}_d \rightarrow \pi^0 \pi^0 $ 
& $\lplp{231}{211}$ & $1.9 \times 10^{-3}$ \\
& $\lplp{331}{311}$ & $5.8 \times 10^{-3}$ \\
\end{tabular}
\end{table}

\begin{table}
\caption{\label{haha2}
The maximum values of
$r_D$ for $CP$ violating $B$ decays with $L$- and $R_p$-violating couplings
assuming $V_{\rm CKM}$ is given by only down-type quark sector mixing.
}
\vspace{0.1 cm}
\begin{tabular}{llcc}
Decay Mode & Sub-quark process & $\phi_{\rm SM}$ & $r_D$ \\
\hline
$\bar{B}_d \rightarrow \psi K_S$ &
$b\rightarrow c\bar{c}s$ & $\beta$ & 0.09 \\
\hline
$\bar{B}_d \rightarrow \phi K_S$ &
$b\rightarrow s\bar{s}s$ & $\beta$ & 2.0 \\
\hline
$\bar{B}_d \rightarrow \pi^0 K_S$ &   
$b\rightarrow u\bar{u}s$, $b\rightarrow d\bar{d}s$ & $\beta$ &  2.8 \\
\hline
$\bar{B}_d \rightarrow D^+ D^-$ &
$b\rightarrow c\bar{c}d$ & $\beta$ & 0.09 \\
\hline
$\bar{B}_d \rightarrow D_{CP} ~\pi^0(\rho^0)$ &
$b\rightarrow c\bar{u}d$, $b\rightarrow u\bar{c}d$ & $\beta$ & 0.06 \\
\hline
$\bar{B}_s \rightarrow \phi K_S$ &
$b\rightarrow s\bar{s}d$ & $\beta$ & 8.0 \\
\hline
$\bar{B}_d \rightarrow \phi \pi^0$ &
$b\rightarrow s\bar{s}d$ & $2\beta$ & 66 \\
\hline
$\bar{B}_d \rightarrow \pi^+ \pi^- $ &
$b\rightarrow u\bar{u}d$ & $\alpha$ & 0.04 \\
\hline
$\bar{B}_d \rightarrow \pi^0 \pi^0 $ &
$b\rightarrow u\bar{u}d$, $b\rightarrow d\bar{d}d$ & $2\beta$ & 3.0 \\
\end{tabular}
\end{table}

\begin{table}
\caption{\label{haha3}
The product combinations which contribute to each decay mode and
the maximum values of
$r_D$ for $CP$ violating $B$ decays with $B$- and $R_p$-violating couplings.
Present constraints on 
the magnitudes of the product combinations are also shown 
\protect\cite{bha,carl}.
}
\vspace{0.1 cm}
\begin{tabular}{llcc}
Decay Mode & Combination & Constraint & $r_D$ \\
\hline
$\bar{B}_d \rightarrow \psi K_S$ & $\lpplpp{212}{213}$ & 
$6.4\times 10^{-3}$ & 12 \\
\hline
$\bar{B}_d \rightarrow \pi^0 K_S$ & $\lpplpp{212}{213}$ &
$6.4\times 10^{-3}$ & 7.2 \\
\hline
$\bar{B}_d \rightarrow D^+ D^-$ & $\lpplpp{212}{223}$ & 
$7.8 \times 10^{-3}$ & 3.2 \\
\hline
$\bar{B}_d \rightarrow D_{CP} ~\pi^0(\rho^0)$ & $\lpplpp{212}{132}$ 
& 1.6 & 3000 \\
\hline
$\bar{B}_s \rightarrow \phi K_S$ & $\lpplpp{212}{223}$ & 
$7.8 \times 10^{-3}$ & 25 \\
\hline
$\bar{B}_d \rightarrow \phi \pi^0$ & $\lpplpp{212}{223}$ &
$7.8 \times 10^{-3}$ & 680 \\
\hline
$\bar{B}_d \rightarrow \pi^+ \pi^- $ & 
$\lpplpp{112}{123}$ & $1.3\times 10^{-6}$ & $1.4\times 10^{-3}$ \\
\hline
$\bar{B}_d \rightarrow \pi^0 \pi^0 $ & $\lpplpp{112}{123}$ 
& $1.3\times 10^{-6}$ & 0.01 \\
\end{tabular}
\end{table}

\end{document}